# Data Requests and Scenarios for Data Design of Unobserved Events in Corona-related Confusion Using TEEDA

Teruaki Hayashi, Nao Uehara, Daisuke Hase, Yukio Ohsawa
School of Engineering,
The University of Tokyo
Tokyo, Japan
hayashi@sys.t.u-tokyo.ac.jp, {nao221997vegas, daisuke00815}@gmail.com, ohsawa@sys.t.u-tokyo.ac.jp

*Abstract*—Due to the global violence of the novel coronavirus, various industries have been affected and the breakdown between systems has been apparent. To understand and overcome the phenomenon related to this unprecedented crisis caused by the coronavirus infectious disease (COVID-19), the importance of data exchange and sharing across fields has gained social attention. In this study, we use the interactive platform called treasuring every encounter of data affairs (TEEDA) to externalize data requests from data users, which is a tool to exchange not only the information on data that can be provided but also the call for data–what data users want and for what purpose. Further, we analyze the characteristics of missing data in the corona-related confusion stemming from both the data requests and the providable data obtained in the workshop. We also create three scenarios for the data design of unobserved events focusing on variables.

## I. INTRODUCTION

Due to the global violence of the novel coronavirus, various industries have been affected and the breakdown between systems has been apparent. To understand and overcome the phenomenon related to this unprecedented crisis caused by the coronavirus infectious disease (COVID-19), the importance of data exchange and sharing across fields has gained social attention. In fact, Johns Hopkins University uses data from the US Centers for Disease Control and Prevention, World Health Organization, and Chinese authorities to visualize the spread of COVID-19 and disseminate information[1]. In addition, local governments and companies, such as the Tokyo Metropolitan Government, are making efforts to disclose data and technology using GitHub [2]. Moreover, the COVID-19 Data Exchange Initiative[3], the pro bono effort launched against the COVID-19 pandemic, has applied their respective expertise, experience, and network to create the largest community of private and public organizations in support of data exchange.

However, the new issue arisen in the corona related co However, the new issue arising in the corona-related confusion is the discussion to determine what types of data are missing. Apparently, cross-disciplinary data sharing and utilization are essential for understanding and controlling unknown phenomena. For this reason, the data published by many institutions are attracting attention, but the intention and background of such data acquisition are often unclear, to the extent that there is insufficient context to grasp the facts and make appropriate decisions. Regarding COVID-19, Silver said, "the number of reported COVID-19 cases is not a very useful indicator of anything unless you also know something about how tests are being conducted" [1]. He warns against looking at statistical data without understanding how and why the data were obtained. Although many international organizations and companies publish some of their data, the data we want are kept fully closed. In other words, it is limited to unilateral information provision from data providers, and there has been almost no discussion about creating data of unobserved events and the methodologies for supporting it.

In this study, we use the platform TEEDA (Treasuring Every Encounter of Data Affairs) [2] to externalize the information from data users. TEEDA is a tool that is specifically developed to exchange not only the information about data that can be provided but also the call for data— what is wanted by data users, and for what purpose. Using TEEDA, we collect data items (data requests and providable data) in the corona-related confusion in the workshop, discuss the characteristics of missing data, and create three scenarios for data design of unobserved events focusing on variables.

The remainder of this paper is organized as follows. In Section 2, we explain the methodology of TEEDA based on the descriptions of data requests and providable data, as well as demonstrate the functions of the platform. In Section 3, we present the experimental details of this study. In Section 4, we discuss the results obtained from our experiment. Finally, we conclude the paper in Section 5.

## II. TEEDA

With the development of data catalogs and portal sites in the data exchange ecosystem, such as data marketplaces, data users have more opportunities to learn about the publications of data holders or providers [3,4]. In the context of corona-related confusion, open platforms for sharing data across crises and organizations, such as the Humanitarian Data Exchange[4] or the World Bank[5], collect and publish many types of data from different domains. However, it is hard to discover information about what types of data the users want, and for what purpose, as this type of detail is not often sufficiently shared. Therefore, data providers are unable to learn what types of data are required, and there is a risk that only those data that do not meet the user requests are provided on the platform. TEEDA facilitates communication and matching between data holders and users by capturing and presenting requests for data (call for data) that users desire in the data exchange ecosystem. The data holders register the information about the data (metadata), and the data users provide information about the purpose and structure of the data in the form of data requests in TEEDA. The collected data items (the data request and providable data) are processed by a matching algorithm and visualized to facilitate data exchange between data holders and users.

---

[1] https://coronavirus.jhu.edu/map.html
[2] https://github.com/tokyo-metropolitan-gov/covid19/blob/development/docs/en/README.md
[3] https://www.covid19-dataexchange.org/
[4] https://data.humdata.org/event/covid-19
[5] https://data.worldbank.org/

Each data request has three description items: data name, variables, and the purpose of data use. The data name is an item to express the data that the users want. Examples include "the rate of self-restraint due to COVID-19" and "behavioral history of those infected with COVID-19." The second description item, variables, is a set of logical data attributes [5,6]. For example, in the case of meteorological data, "area name," "maximum temperature," "minimum temperature," "average temperature," or "date" are the variables. The third description item captures an expression of the purpose of data use. In this study, this item will be useful for understanding what types of data and variables are needed and for what purpose in regard to corona-related confusion. As an existing part of TEEDA, the records of providable data already have some description items in an element known as the "data jacket" (DJ). A DJ is a framework for summarizing data information while keeping the data itself confidential [7]. The summary information of the data includes explanatory text about the data. A DJ enables an understanding of the types of data that exist on different platforms and the information included in the data, even if the contents of the data cannot be made public. In the TEEDA format for providable data, we used data name, data outline, variables, types, formats, and sharing conditions of data. The sharing condition is the list of terms and conditions imposed by data providers to exchange data with, or provide data to, other parties.

We used variables to examine the relationships and matching possibilities of data requests and providable data, based on the assumption that the completeness of the variables is the condition for data users. Therefore, we can represent the relationships between data requests and providable data in the network format, where the data items are the nodes and the links are established when the data items have common variables. Figure 1 shows the TEEDA interface and the network of data items input in the experiment (explained in detail in the next section). The green nodes represent the data requests, and the orange nodes are the providable data. To encourage data users and providers to understand the relationships between their own data items as well as others', links are also established between data requests and between providable data. TEEDA runs on a Web browser, and the input data items are reflected on the screens of other users in real-time.

TEEDA will automatically highlight neighboring nodes when browsing the data item details of a clicked node. In addition, there is an interface between the dashboard and the toolbox shown on the right side of Fig. 1, where the input data items and variables are displayed. The network layout can be changed manually by dragging and dropping.

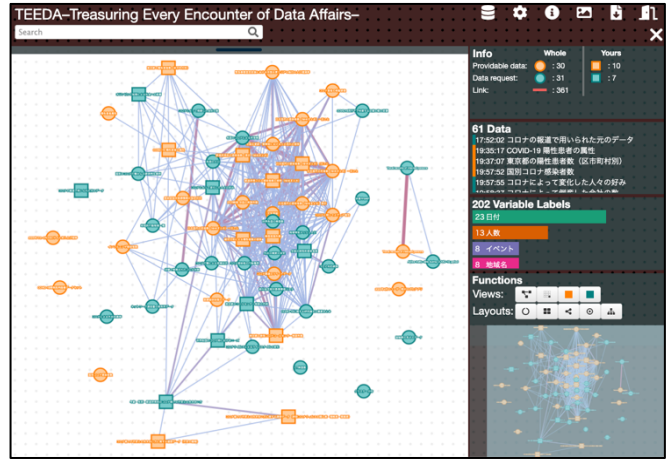

Fig. 1. TEEDA interface and collected data items (data requests and providable data) with their relationships.

In this study, we focus on the function of call for data of TEEDA and externalize the needs for data in corona-related confusion as data requests. In addition, we analyze the degree to which data requests are satisfied by comparing with the data that have been released during corona-related confusion as providable data. Subsequently, we propose and discuss three scenarios of data consisting of what types of variables should be newly designed and acquired.

### III. EXPERIMENTAL DETAILS

The aim of the experiment was to understand the characteristics of data requests and providable data in the corona-related confusion and create scenarios for new data design of unobserved events focusing on variables. The experiment involved 14 men and women (students and professionals) 20 years and older. Initially, they were taught how to use TEEDA for approximately 15 min. Subsequently, participants input the information on the data requests and the providable data about corona-related confusion on TEEDA for 45 min via discussion with other participants.

According to the specifications of the TEEDA platform, when data requests are entered, only the description items of 'data name' and 'variables' are required. The 'purpose of data use' description item is optional. All items are written in natural language, and there is no upper limit to the number of words entered. When entering providable data, the data name and variables are required, and the data outline is optional. As before, all items are written in natural language with no upper limit to the word count. There are nine recognized datatypes: "time series," "numerical value," "text," "table," "image," "graph," "movie," "sound," and "other,". TEEDA can deal with nine file formats: "CSV," "txt," "RDB," "markup," "RDF," "weka," "shape," "PDF," and "other." Users can select multiple checkboxes for these items because some data have multiple data types and are provided in several file formats. By contrast, users select one sharing condition with a radio button from seven predefined types: "generally shareable," "conditions/negotiations are required," "shareable within a limited range," "non-shareable," "shareable by purchased," "not yet decided," and "other conditions."

Note that the providable data externalized in the workshop did not necessarily cover all the available data in corona-related confusion. Since the experiment was conducted on June 15th, 2020, it should be noted that the results obtained and the attributes of some data may have changed after this paper was published. In addition, the input information about data in the workshop was written in English and Japanese, and in the analysis, we unified them into English.

## IV. RESULT AND DISCUSSION

### A. Characteristics of Data Request and Providable Data

Sixty-one data items—divided approximately evenly into 33 data requests and 28 providable data entries—were input during the workshop. Tables I and II show examples of data requests and providable data, respectively. First, we discuss what types of data and for what purposes they are required in corona-related confusion. In the data requests, a lot of data are necessary for understanding the measures to prevent infection by social distancing or quarantine, such as "Behavioral history of those infected with COVID-19" and "Measures against COVID-19 implemented at stores." In addition, there are many lifestyle-related data including data for managing anxiety such as "Coping with anxiety during COVID-19 pandemic by age, sex, and prefecture," "Changes in the lifestyle caused by COVID-19," or "People's preference changed after the COVID-19 pandemic." By contrast, to recognize the facts, there were the need for new statistical data for supplementing the published data, such as "Needs of countries in the world during COVID-19 pandemic" and "Number of tests in countries around the world." Typical purposes of these data were "To analyze the situations of different countries because it is hard to compare with current published data."

Most of the providable data were statistical data on the attributes of infected persons, such as "Number of COVID-19 cases by country," or "Number of positive cases in Tokyo Metropolis (by city)." These data were mainly published by governments and international institutions. In addition to statistical data, there were publicly available data for academic purposes, such as "Image datasets for COVID-19 related physicians", as well as survey data provided by the investigation company, such as "A survey on coping with anxiety during COVID-19 pandemic" related to staying home or working from home.

TABLE I. EXAMPLE OF DATA REQUEST

| Item | Content |
|---|---|
| Data name | Needs of countries in the world during COVID-19 pandemic |
| Variables | Country, needs, product name, service name, reason, age, age group, address |
| Purpose of data use | There was hoarding and a toilet paper shortage. We must clarify what products were really needed and lacking in practice. |

TABLE II. EXAMPLE OF PROVIDABLE DATA

| Item | Content |
|---|---|
| Data name | Trends in the number of positive cases by date of confirmation |
| Variables | Total number of cases, daily number of cases, date |
| Data outline | Open data provided by the Tokyo Metropolitan Government. For more accurate analysis of the trends of new COVID-19 cases, information on new COVID-19 cases reported from public health centers was organized by the date of confirmation by physicians through PCR testing. |
| Types | Time series, number, table, image |
| Formats | CSV, others |
| Sharing conditions | Generally shareable |

Figure 2(a) shows a comparison of the providable data collected using TEEDA in the past and the data in corona-related confusion by sharing conditions. We used 234 cases whose sharing conditions are described in the past providable data. Consequently, the proportion of shareable data provided in corona-related confusion was about 90%, whereas the ratio before Corona was only 35%. In other words, a large amount of 'generally shareable' data are externalized as being relevant to corona-related confusion. It is known that the ratio of generally shareable data in the data exchange platform is about 50%[8], and institutions and companies may tend to be more open with their data related to problems with high public interest, such as corona-related confusion.

By contrast, the comparisons of data types and formats are shown in Fig. 2(b) and (c). Note that one piece of data can have multiple types and formats. The ratios of data types are almost the same for "time series," "numerical value," "text," "table," and "image," but it can be seen that the proportion of the "graph" under the corona-related confusion is significantly larger than that before corona. This is because the data related to this topic are often provided in a graph format so that even a person who is not a specialist in the data can read it and understand the trend and the situation of the number of infected persons. In fact, 13 of the 14 data that have "time series" also have the data type "graph." In addition, tabular data are also an excellent format for reading and comparing as well as the graph data, and many data on the number of infected persons by prefectures include the type "table."

By contrast, regarding data formats, there are many data in CSV and RDB formats, which easily handle time-series data in the tabular form, and TXT of the language corpus in the data before corona. However, for the data since corona, "other" format is the most frequent with 11 cases. All 11 of the "other" data are the image format (such as jpeg or png), including "image" or "graph" for the data types, also peculiar to the corona situation. Image formats are good at visually conveying information to the public. By contrast, it seems that too much emphasis is placed on just communicating the information because the data that allow secondary use, such as CSV, are seen less often since the pandemic began. The data provided in PDF, which is human-friendly but has poor machine readability, also exists in a certain proportion under the corona situation, and it is required to provide data in a format that makes it easy for secondary use.

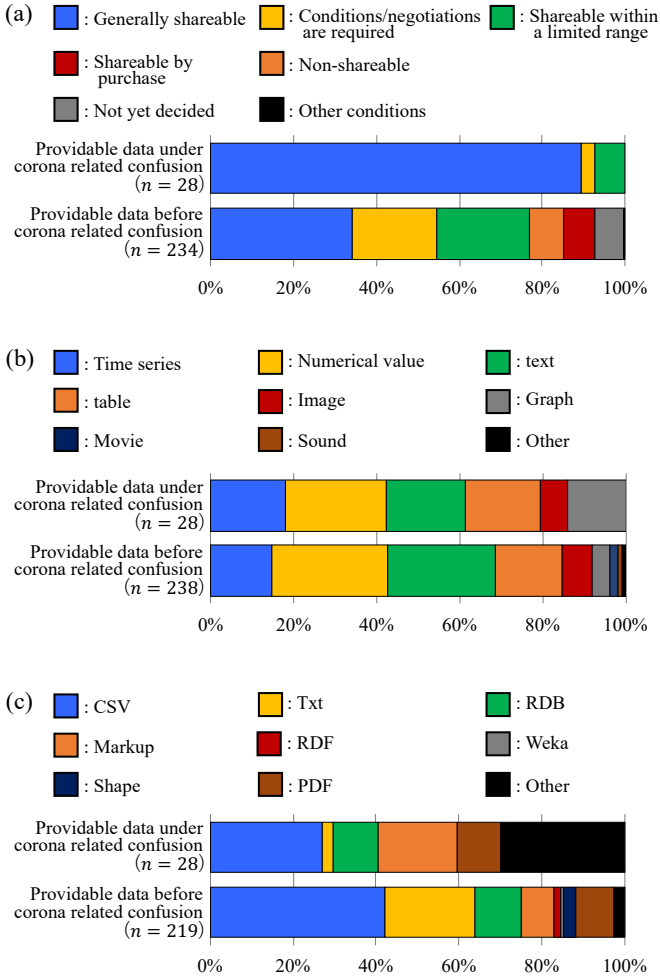

Fig. 2. Comparisons of providable data under/before corona-related confusion by (a) sharing conditions, (b) data types, and (c) data formats.

TABLE III. CHARACTERISTIC VALUES OF DATA REQUESTS AND PROVIDABLE DATA WITH VARIABLES

|  | All data | Data request | Providable data |
|---|---|---|---|
| No. of data items | 61 | 33 | 28 |
| No. of variables | 315 | 156 | 159 |
| Types of variables | 193 | 100 | 113 |
| Average no. of variables in each data item | 5.16 | 4.73 | 5.68 |
| Maximum no. of variables in each data item | 18 | 8 | 18 |
| Minimum no. of variables in each data item | 2 | 2 | 2 |

Table III presents the details of the input data items. The types of variables in the providable data were 113, which is slightly larger than those of data requests. In addition, the average number of variables is larger at 5.68 in providable data than that of data requests at 4.73, which varies from 2 to 18. In contrast, the number of data that the data users want to obtain is as large as 33, but both the average number and the types of variables are less than those of providable data. These results suggest that data users may not need the data composed of many variables and do not require variables as diverse as the providable data. This is an important point when considering data design scenarios, and will be discussed in detail in the next subsection. As for the frequency of variables, 79 of 100 types of variables appear only once in the data requests and 94 of 113 types of providable data appeared once. In other words, the frequency of appearance of most variables is approximately once in both data requests and providable data. In a previous study, the frequency distributions of the variables of both data requests and providable data show power distributions [8]. Although the number of samples in this experiment is small, it is considered that both data requests and providable data are composed of a variety of variables with low frequency.

Next, we compared the details of the types of variables in the data items that appeared in corona-related confusion. Figure 2 shows the top-15 variables of (a) all data items, (b) data requests, and (c) providable data. The variable "date" appeared the most; while "number", "prefecture name", and "area name" of the patients consistently occupied the top ranks. When discussing the variables, it is debatable whether to use well-defined schemata or natural language concepts. Studies on ontology matching [9,10] or ontology-based data access [11,12] have defined schemata for heterogeneous data integration. Because corona-related confusion is an unprecedented crisis, the types of data that are providable or needed remain unclear. To allow diverse data with a variety of variables to understand and make decisions during the crisis, although "address" and "location" are almost synonymous, in this study, we did not unify the notation fluctuation of variables.

As we explained, most of the providable data were statistical data concerning the attributes of infected persons (patient's place of residence, city name, the degree of seriousness) along with other variables, such as the number of polymerase chain reaction (PCR) tests or the event names, which are orderable in a time series by the inclusion of the variable "date." By contrast, some data such as "Image datasets for COVID-19 related physicians" and "A survey on coping with anxiety during COVID-19 pandemic" do not have "date", and are unique compared with other providable data. In the data requests, "area name" and "address" whose granularities are higher than "prefecture name" or "city name" appeared frequently, which are not included in the statistical data in the providable data. These variables are included in the data such as "Measures against COVID-19 implemented at stores" and "Number of cases by hospital in Japan", and the reasons why these data were required are "To use them as best practices." The rate of infection to the number of healthcare workers is investigated to identify the risk of overwhelming hospitals," which were desired for grasping the current situation of the corona and taking measures. By contrast, not only the high granularity variables such as "address" or "area name" but also it is interesting that there is a large-meshed variable such as "country name." As described above, it is a central variable for accurately understanding the global situation of COVID-19 rather than individual decision making. Furthermore, looking at the breakdown of the types of variables, there were only 20 types of variables common between data requests and providable data. This result means that the providable data do not contain enough variables for the data that users want to obtain. It can be said that there is a big mismatch here.

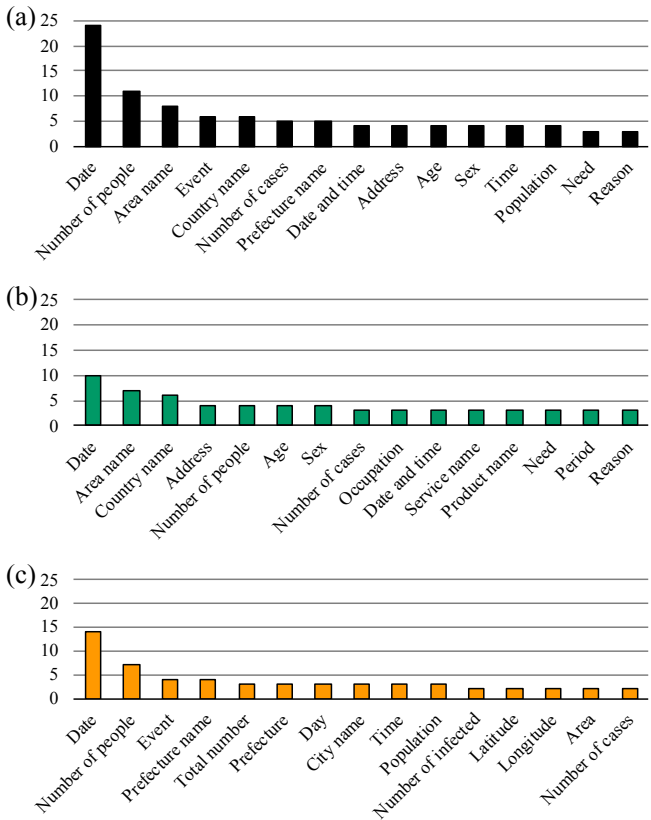

Fig. 3. The top 15 variables (a) in the data requests and providable data, (b) in the data requests, and (c) in the providable data.

TABLE IV. NUMBER OF CATEGORIZED DATA REQUESTS WITH EXAMPLES

| Category | # of data requests | Examples |
|---|---|---|
| Phenomenon Understanding | 15 | • Needs of countries in the world during COVID-19 pandemic<br>• Number of tests in countries around the world<br>• Business status impacted by COVID-19<br>• MeSH population by time of day |
| Individual decision-making | 7 | • Recommended frequency of going out<br>• Data on the number of contacts<br>• Behavioral history of those infected with COVID-19<br>• Travel records of infected persons |
| Organizational decision-making | 11 | • Impact of the postponement of the 2020 Olympics on society<br>• The rate of self-restraint due to COVID-19<br>• Coping with anxiety during COVID-19 pandemic by age, sex, and prefecture<br>• People's preference changed after the COVID-19 pandemic |

## B. What kinds of data should we design?

The results in the last subsection suggest that although the data provided under the corona-related confusion are varied in the types of variables, there are relatively few data that satisfy the data requests, leading to a mismatch. What types of data need to be newly designed and obtained in corona-related confusion? There are various purposes for using data in data requests, and we categorized them into the following three types:

- Phenomenon understanding: for verifying the facts that have affected society, such as business or medical fields
- Individual decision-making: to obtain evidence for making decisions in one's life, such as going out or staying home
- Organizational decision-making: for learning the social demands and changes in the post-corona society and formulating business and organizational guidelines

Table IV shows the number of categorized data requests with examples. We analyzed the types of variables in each category and discussed possible scenarios for data acquisition.

### 1) Scenario 1 (Phenomenon understanding)

Fifteen data requests are categorized as "phenomenon understanding," which was the most numerous compared with the other two categories. The variables "country names (6 times)," "area names (4 times)," and "date (4 times)" appeared frequently, and the requests contained both the detailed and global variables lacking in the data provided by local governments and institutions. In particular, many variables for understanding what types of needs from which ages in corona-related confusion have received much attention. In understanding the phenomena, it is better to acquire the data with missing variables included in the data requests, using "date," "area name," and "address." In addition, although we understood that many companies were affected by the COVID-19 pandemic, there are few data on the types of companies in which industries were affected. It is considered important to collect data to understand the kinds of impacts with the variables "area name," "type of business."

### 2) Scenario 2 (Individual decision-making)

There are seven data requests related to "individual decision-making," least numerous compared with the other two categories. Among the variables, "date (3 times)" is the most common, "address (twice)" is the second most common, and all others appeared only once, with the ratio of variables that appear only once being the highest among the three categories. Variables such as "recommended frequency of going out," "acceptance of COVID-19 patients," "item people touch," and "number of people touching it per day" are the unique events that have not been observed yet. It is difficult to extract common interests because of the diversity of needs in individual decision-making, but there seems to be a need to obtain data that are deeply related to our lives, such as the data on the number of contacts or the infected information in the area where we live.

### 3) Scenario 3 (Organizational decision-making)

In the decision-making of the organization, "sex (4 times)," "age (4 times)," "area name (3 times)," and "date (3 times)" appeared frequently. Based on these common variables, there are the data including variables to try to change their business policies such as "increased activity due to self-restraint," "whether to continue it or not after self-restraint life," and the variables to create new businesses such as "type of anxiety" or "consultation content." For companies and institutions to adapt to social changes in people's lives in the wake of the corona pandemic, we consider there to be a need for extended statistical questionnaire data beyond that, which has not been widely provided yet.

Although there were differences in the types of variables required for each category, it can be said that "date" is the central variable in the data design of unobserved events in all categories. In particular, "date" plays an important role not only as statistical data but also as it captures people's interests and business conditions that change from moment to moment in corona-related confusion. Moreover, it is notable that the number of variables in each data request are few and it suggests that users want the data with just those variables specialized to their own interests, which is hardly included in the providable data. In addition, from the analysis of providable data, it can be said that data formats that are not only human-friendly, such as PDF or images, but also machine-readable and easy for secondary use, such as CSV or JSON, are strongly required.

## V. CONCLUSION

### A. Summary

In this study, to discuss the data design of unobserved events in corona-related confusion, we used TEEDA to externalize the information about data items from data users and data providers and analyzed their characteristics. Via experiments, we found different structures across data requests and providable data and the large mismatch between them. Based on the discussion, we created three possible scenarios for data design, focusing on variables in data requests divided into three categories: phenomenon understanding, individual decision-making, and organizational decision-making. In our future work, we will obtain data according to these scenarios and verify via demonstration experiments whether the results meet the needs of data users in the society.

### B. Future Work

In this study, we obtained data requests and providable data from the participants in the form of the workshop, but it was not possible to cover all the available data provided in the actual corona pandemic. There are more data and various variables in the world. To find out more information about providable data, it is important to collect them differently and discuss their characteristics.

Moreover, from the viewpoint of data design, the variables contained in data other than corona-related data are also considered useful. Variable Quest (VQ) is an algorithm with the knowledge base for estimating sets of variables of unknown events from data outlines [13]. Using the knowledge base of VQ for external information about data and variables, it may be possible to construct data for unobserved events. In addition, since major file exchange formats such as JSON were not yet supported by TEEDA, these specification changes are possible considerations for the future.

ACKNOWLEDGMENT

This study was supported by JSPS KAKENHI (JP20H02384), the "Startup Research Program for Post-Corona Society" of Academic Strategy Office, School of Engineering, the University of Tokyo, and the Artificial Intelligence Research Promotion Foundation. We wish to thank Editage (www.editage.jp) for providing English language editing.